# ResAtom System: Protein and Ligand Affinity Prediction Model Based on Deep Learning

Yeji Wang[a], Shuo Wu[a], Yanwen Duan[a,b,c], and Yong Huang[a,c,*]

[a]Xiangya International Academy of Translational Medicine, Central South University, Changsha, Hunan, 410013, China

[b]Hunan Engineering Research Center of Combinatorial Biosynthesis and Natural Product Drug Discover, Changsha, Hunan, 410011, China

[c]National Engineering Research Center of Combinatorial Biosynthesis for Drug Discovery, Changsha, Hunan, 410011, China

*Corresponding author.

E-mail address: jonghuang@csu.edu.cn (Y. Huang)

**Abstract:**

**Motivation:** Protein-ligand affinity prediction is an important part of structure-based drug design. It includes molecular docking and affinity prediction. Although molecular dynamics can predict affinity with high accuracy at present, it is not suitable for large-scale virtual screening. The existing affinity prediction and evaluation functions based on deep learning mostly rely on experimentally-determined conformations.

**Results:** We build a predictive model of protein-ligand affinity through the ResNet neural network with added attention mechanism. The resulting ResAtom-Score model achieves Pearson's correlation coefficient R = 0.833 on the CASF-2016 benchmark test set. At the same time, we evaluated the performance of a variety of existing scoring functions in combination with ResAtom-Score in the absence of experimentally-determined conformations. The results show that the use of $\Delta_{Vina}RF_{20}$ in combination with ResAtom-Score can achieve affinity prediction close to scoring functions in the presence of experimentally-determined conformations. These results suggest that ResAtom system may be used for *in silico* screening of small molecule ligands with target proteins in the future.

**Availability:** https://github.com/wyji001/ResAtom

## 1. Introduction

The rapid discovery of drug hits with superior biological activity, as well as low toxicity and few side effects still faces huge challenges. It is not only time-consuming but very expensive from the discovery of these potential hits to their final approval to treat human diseases by drug-regulatory agencies world-wide (Paul et al., 2010). Although multi-target drugs and many transforming therapeutics have made tremendous progress, small molecule drugs against a single protein target are still the primary choice for the treatment of many diseases (Chen et al., 2016; Tonge, 2018). However, it is not suitable for large-scale and high-throughput screening, due to the high cost associated with those experimental measurement of protein and ligand affinity, (Yang et al., 2019). Especially for rare diseases, it is necessary to accelerate drug discovery and development methods without incurring high costs due to parallel development (Stecula et al., 2020). Therefore, accurate and rapid determination of the affinity between proteins and their potential ligands plays a vital role in the search for new drug hits (Myers and Baker, 2001).

To reduce research costs and improve research efficiency, a variety of studies have been focused on the development of efficient and accurate *in silico* screening methods of protein targets and large small molecule libraries to discover potential ligands with high affinity (Mathai and Kirchmair, 2020). To improve screening efficiency, people have developed a variety of algorisms to calculate protein-ligand affinity (Cao and Li, 2014; Leelananda and Lindert, 2016; Rezaei et al., 2020).

After decades of accumulation, a large amount of structure and activity data of proteins and their ligands have been accumulated, resulting in many data sets with a high level of confidence between protein-ligand interactions and their *in vitro* binding affinity (Mysinger et al., 2012; Wang et al., 2005). These invaluable and properly compiled large data sets make it possible to build scoring models to study new protein-ligand interactions and to predict their affinity of potential ligands to the target proteins, using certain deep learning algorithms.

In recent years, several groups used machine learning or deep learning to predict the affinity of proteins and their potential ligands (**Figure 1**). Comparing with traditional scoring functions, deep learning would learn higher-dimensional interaction features of protein-ligands and implement more complex functions to reflect their intrinsic interaction and affinity relationships. For example, Durrant et al. established a neural network scoring model with two hidden layers (Durrant and McCammon, 2010; Durrant and McCammon, 2011), while Jiménez, Marta, or Ragoza and co-workers used 3D convolutional neural networks (3D-CNN) with different frameworks to build more accurate scoring models (Jimenez et al., 2017; Jimenez et al., 2018; Ragoza et al., 2017). Recently, Zheng et al. converted the 3D information of protein-ligands into 2D data and a 2D-ResNet neural network was used to build a scoring model (Zhang et al., 2019). Kwon et al. used the ResNet network to build a scoring model and used an ensemble method to improve its performance (Kwon et al., 2020). A very recent achievement is a relatively lightweight neural network built on ShuffleNet, with a better performance (Rezaei et al., 2020). McNutt et al. have recently also used convolutional neural networks to optimize molecular docking (McNutt et al., 2021).

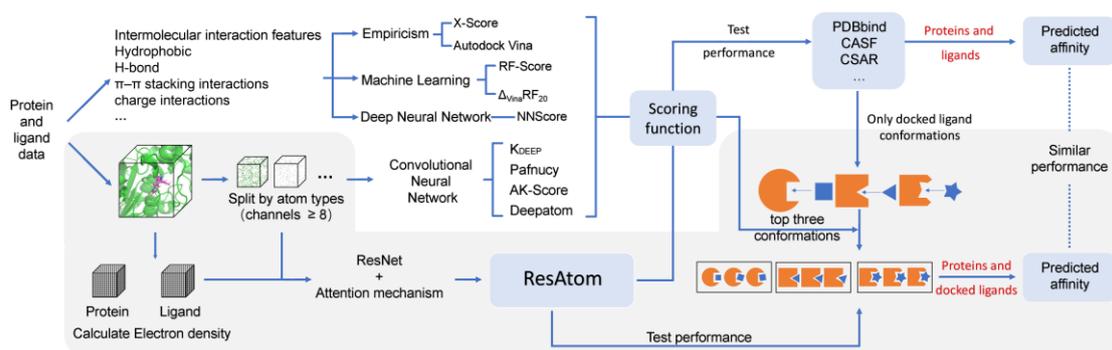

**Figure 1**. The summary of recent scoring functions and affinity prediction algorithms for protein-ligand affinity prediction and the ResAtom algorithm developed in the current study. Autodock Vina (Trott and Olson, 2010) is used to generate protein-ligand conformations in ResAtom, and after certain conformations were filtered by existing scoring functions, a deep learning-based model is used to predict protein-ligand affinity.

Thanks to the development of cryo-electron microscopy, the tremendous progress in protein structure studies have enabled the ever-close understanding of the role of protein-ligand interaction in life (Cheng, 2015; Nakane et al., 2020; Yip et al., 2020). However, the vast human proteome would make the discovery of ligands towards every protein a dauting task, and understanding how they interact would be even more challenging (Schreiber, 2005). To our knowledge, only experimentally measured protein-ligand conformations have been used to evaluate most models previously, while their application towards protein-ligand affinity prediction without experimentally pre-determined structure information is rarely discussed.

In this study, we thus establish and evaluate a new deep learning algorithm named ResAtom to predict protein and ligand affinity, both in the presence and absence of experimentally determined protein-ligand structures (**Figure 1**). The resulting ResAtom-Score model achieved Pearson's correlation coefficient R = 0.833 on the CASF-2016 benchmark test set, showing better performance than $\Delta_{Vina}RF_{20}$-based machine learning or deep learning-based AK-Score algorithms (Kwon et al., 2020; Wang and Zhang, 2017). In the absence of the experimentally determined ligand conformation, the combination of ResAtom-Score and $\Delta_{Vina}RF_{20}$ can still accurately predict protein-ligand affinity with R = 0.828, using the CASF-2016 benchmark test set, which is superior to other tested deep-learning models. Taken together, ResAtom may be conveniently adapted to predict those protein-small molecule affinities, when the detailed protein-ligand structural interactions are not known.

## 2 Material and methods

In this study, we established a protein-ligand affinity evaluation system using ResNet-based deep learning framework as shown in **Figure 1**.

### 2.1 Build ResAtom-Score

2.1.1 Preparation of dataset

We chose the PDBbind database as the preferred dataset, since it contains complete information of proteins and their ligands, such as the three-dimensional structure and experimentally-determined affinity information (Liu et al., 2017). The PDBbind dataset is divided into general and refined set, which contains 12 800 and 4 852 data sets, respectively. To keep as much data as possible, we did not filter the PDBbind database based on protein resolution. In general, we adapted the following rules to preprocess the PDBbind database (**Figure 2**): 1. The entries containing peptide ligands were removed. 2. The entries containing covalently-bound ligands were removed. 3. The entries containing incomplete ligands were removed. 4. Oversized proteins were removed. After the above processes, a total of 15 038 structures with protein-ligands were trained and both training set and validation set are listed in **SI-1**.

We next used Rdkit and OpenBabel (O'Boyle et al., 2011) to process proteins and ligands in the dataset, while protein and ligand data using in training and validating are independent of test set. We used the HTMD Python framework to convert the structural information into a fixed grid centered on the geometric center of the ligand (Doerr et al., 2016). Each side of the grid is 35 A, centered on the ligand, and the resolution is 1 A. We divided the atoms in the protein and ligand into eight types and calculate the voxelization information separately. Then we used the promolecular method in Multiwfn to calculate the electron density of proteins and ligands (Lu and Chen, 2012). We integrated the calculation results and randomly divided the processed data into training sets and validation sets by a ratio of 8 : 2 in each training for each single model. This dataset is independent of benchmark tests, including CASF-2016 and CSAR-HiQ.

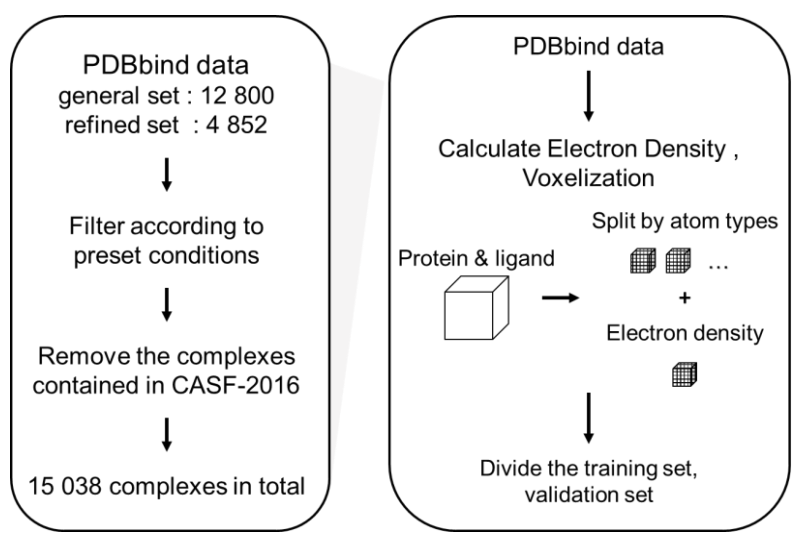

**Figure 2.** The process of voxelization of PDBbind data

Finally, each protein and ligand contain 18 channels. The atom types of the first 16 channels are the same as Jiménez et al. (Jimenez et al., 2018), including hydrophobic, aromatic, hydrogen bond acceptor, hydrogen bond donor, positive ionizable, negative ionizable, metallic, excluded volume. The last two channels are the electron densities of proteins and ligands, respectively.

2.1.2 Network architecture of ResAtom-Score

The ResNet architecture improves learning performance by increasing the depth of the network and alleviates the problem of a relatively small training dataset (He et al., 2016). Therefore, we chose PyTorch to build the ResNet network with the addition of an attention mechanism (Woo et al., 2018). This framework can build deeper models while reducing the difficulty of online learning. The structure of the 3D attention-based residual neural network is depicted in **Figure 3**, and we named it ResAtom.

ResAtom consists of a convolutional layer, an attention block, sixteen basic residual blocks, and a fully connected layer. Each residual block consists of two convolutional layers, two batch normalization layers, and a rectified linear unit (ReLU) (Nair and Hinton, 2010). We reasoned that using the average pooling layer instead of the maximum pooling layer would reduce the model's sensitivity to slight changes in conformational root mean squared deviation (RMSD). The attention block includes two parts, e.g., spatial attention and channel attention. Spatial attention would generate a model with stronger focus on combining relevant information, while channel attention would result an intelligent model knowing what kind of information is more important. Therefore, the introduced attention mechanism in ResAtom-Score may improve its representation ability and focus on important features instead of trivial parts.

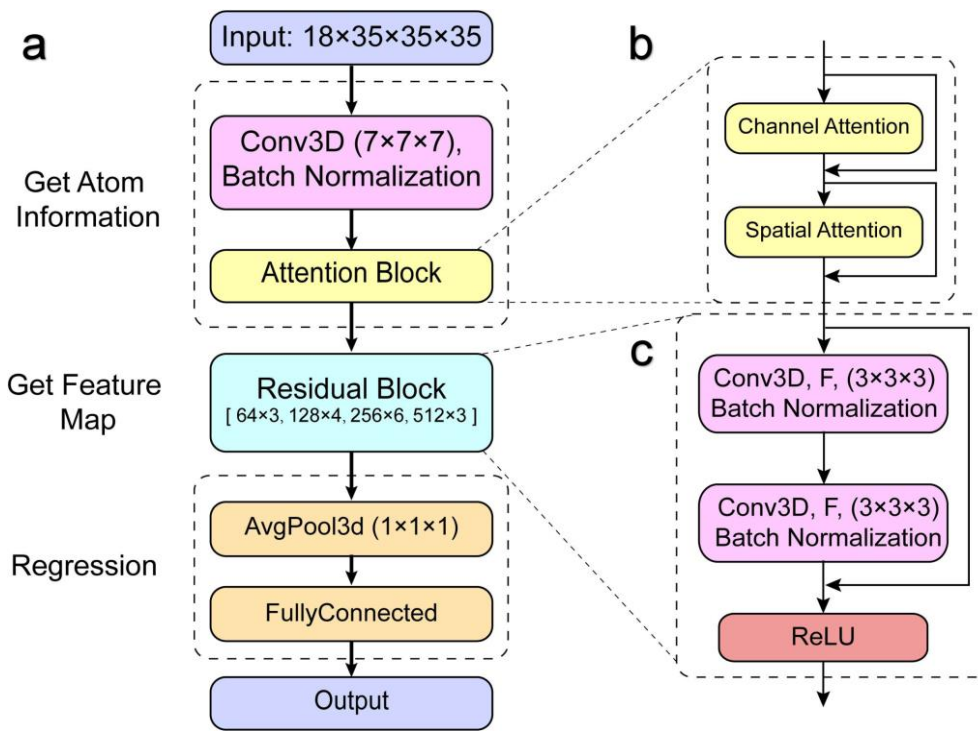

**Figure 3.** (a) The architecture is mainly composed of three modules: preprocessing module, feature map extraction module, and regression module. (b) The attention block includes two parts: spatial attention and channel attention. (c) Each residual module consists of two convolution blocks, two batch normalization blocks, and an activation function (ReLU).

2.1.3 Training and testing

To start training the processed training sets, we choose Torchio to randomly translate and rotate the input datasets in various dimensions (Pérez-García et al., 2020). During the training process, different hyperparameters, including learning rate, optimizer, and architecture, were used to evaluate the impact on ResAtom.

When testing, each protein and ligand were randomly translated and rotated for five times, and the average of the predicted affinity was used. We selected the CASF-2016 benchmark test set (Li et al., 2014) to evaluate the performance of ResAtom-Score. The CSAR-HiQ set (Smith et al., 2011) was also used to perform additional tests on ResAtom, in which there are only 75 data sets left after those data in the CSAR-HiQ set were removed due to overlapping with those dataset in our training sets from PDBbind database. The list of the remaining 75 complexes is presented in the **Supporting Information**.

We next studied the affinity prediction performance by integration of existing scoring functions with ResAtom-Dock. We used a few scoring functions to score potential conformations of each ligand and the top three conformations were picked. The ResAtom-Score model was then used to predict the affinity of the top three conformations, and the average of these conformations was used as the final predicted affinity. The Pearson correlation coefficient (R) was used to evaluate the performance of ResAtom-Score.

**2.2 Build ResAtom-Dock**

We also used similar deep learning architecture with ResAtom-Score to train a conformational screening model named ResAtom-Dock. Unlike ResAtom-Score, the fully connected layer of ResAtom-Dock was divided into five outputs (grouped by their respective RMSD values compared with the experimental conformation from 0, 0 – 2, 2 – 6, 6 – 10, > 10), and an activation function SoftMax was added at the end (Bridle, 1990). The popular Autodock vina is used for molecular docking of proteins and ligands in the PDBbind refined set. Several conformations (~10) were generated for each ligand. The docking results of ligands were classified according to their RMSD values and imported into ResAtom-Dock for training. The hyperparameters are the same with ResAtom-Score.

**2.3 CASF-2016 benchmark test set preprocessing to test the performance of ResAtom-Score with selected scoring functions.**

In most cases, people cannot obtain the binding conformation of protein and ligand and some of these interactions may be obtained through simulation, which puts higher requirements on the performance of the scoring function. Therefore, we examined the performance of different scoring functions when used alone or in combination in the absence of experimentally-determined conformations.

To ensure reproducible results, we only used docked ligand conformations provided by CASF-2016. We rank all conformations of each ligand according to different scoring functions and use the scoring function itself or other scoring functions to predict the top three conformations. Finally, the average value of the score is used as the final predicted value. We choose Pearson's correlation coefficient and RMSE to evaluate the results.

**2.4 Implementation**

We used the docked conformations of 285 ligands provided in CASF-2016 benchmark test set to evaluate the performance of ResAtom system. In order to predict the affinity of potential ligands to a given protein using ResAtom system, Autodock Vina is first used to generate multiple docking conformations of these ligands to the target protein through iterative docking. Next, ResAtom-Dock is used select the top three conformations of the ligands. Finally, ResAtom-Score is used to predict the binding affinity of the selected conformations, and the average of their affinity is taken as the final output. Alternatively, other scoring functions, such as X-Score or $\Delta_{Vina}RF_{20}$, can be used to obtain the respective top three conformations of the ligands, followed by running the affinity evaluation script of ResAtom-Score to obtain the predicted affinity of each conformation.

## 3. Results and discussion

### 3.1 ResAtom-Score model performance

To obtain the ResAtom-Score model with more accuracy, we have systematically optimized the hyperparameters during the model training process. We used the correlation coefficient (R) to evaluate the performance of the network. The influence of different learning rates and optimizers on the trained model is shown in **Table S1**. The influence of different model architecture and different learning rate decline strategies on the model are shown in **Table S2**. In the end, we chose Adam (Kingma and Ba, 2014) as the optimizer with a learning rate of 0.001, with a batch size of 256. Mean-square error (MSE) was used as the loss function, while CosineAnnealingLR was used to dynamically adjust the learning rate (Loshchilov and Hutter, 2016).

Since CASF-2016 provides many pre-predicted results of existing scoring functions, we therefore opt to use the same test set to compare ResAtom-Score with existing scoring functions. The results by comparing of different scoring functions with ResAtom-Score is shown in **Figure 4a**.

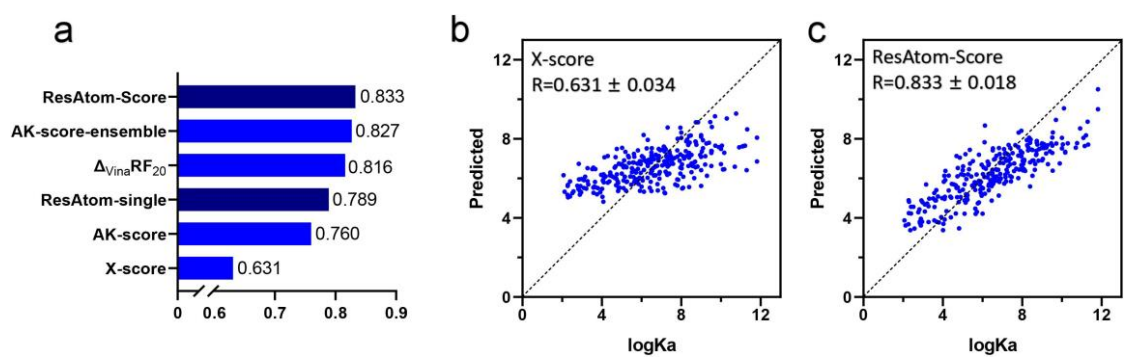

**Figure 4**. The CASF-2016 benchmark test of ResAtom-Score and selected existing scoring functions. (a) The scoring power of ResAtom-Score and existing scoring function (Kwon et al., 2020; Su et al., 2019). (b) The prediction of X-score. (c) The prediction of ResAtom-Score. The Pearson correlation coefficient value of the predictions and their uncertainties are estimated using the bootstrapping analysis.

A single model of ResAtom achieved a Pearson correlation coefficient R = 0.789 for a single model, which exceeded that of the AK-Score (R = 0.760), with similar deep learning architecture. Unique to ResAtom-Score is that the additional electron density of proteins and ligands, as well as dynamic changes in the learning rate and an expanded depth due to use of ResNet-34 framework. We further integrated nine models to result in ResAtom-Score with the correlation coefficient of 0.833, which however is only slightly better than AK-Score-ensemble (R = 0.827), which integrated over 20 single AK-Score models. Using CASF-2016 benchmark test, the distribution of experimental and predicted values of ResAtom-Score and X-score are shown in **Figures 4b and 4c** and ResAtom-Score shows a stronger correlation with the experimentally measured affinity, while X-score, one of the scoring functions of commonly used docking software has a large deviation. When more single ResAtom models were integrated, the performance of the ensembled model declined rapidly.

To evaluate the application capability of ResAtom-Score, we selected the CSAR-HiQ set as an additional test set. At the same time, we compared ResAtom-Score with Cyscore (Cao and Li, 2014) based on empiricism and RF-score (Wójcikowski et al., 2017) based on machine learning (**Table 2**). The predicted root mean squared error (RMSE) of ResAtom-Score is 1.73, and the correlation coefficient between the experimental value and the predicted value is 0.66, which is lower than the CASF-2016 benchmark test set result. Compared with other scoring functions, ResAtom-Score still has certain advantages compared with other scoring functions, despite that there is a certain decline in prediction performance when evaluated using CSAR-HiQ sets. For example, ResAtom-Score shows an improved prediction of logKa in the range of 2 to 10, while Cyscore exhibits a larger deviation (**Figure 5**).

**Table 2. The performance of scoring models using CSAR-HiQ set.**

| Score Model Name | R | RMSE | MAE |
|---|---|---|---|
| RF-score | 0.52 | 2.19 | 1.72 |
| Cyscore | 0.63 | 2.74 | 2.3 |
| ResAtom-Score | **0.66** | **1.73** | **1.42** |

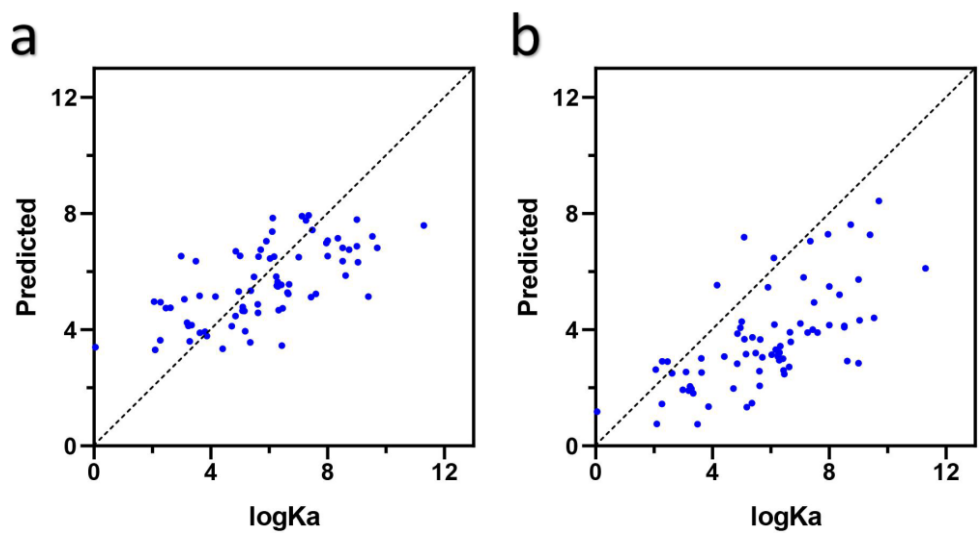

**Figure 5**. Binding affinity prediction results of protein-ligand complexes that are in CSAR-HiQ set but not used for training. (a) ResAtom-Score (b) Cyscore

### 3.3 ResAtom-Dock performance

The performance of our conformation screening model ResAtom-Dock and representative models are shown in **Table S3**. ResAtom-Dock is only better than Autodock Vina, while it is far inferior to the empirical scoring function X-SCORE and the machine learning-based scoring function $\Delta_{Vina}RF_{20}$. We speculated that the reason is that the ligand conformations (~10 for each ligand) in the training sets are diverse, and many of them are in fact distal to the experimentally determined ones. For example, the RMSD distributions of > 80% ligand conformations obtained through Autodock Vina are above 4 (**Figure S2**), which would lead to uneven distribution of the training sets and eventually affect the performance of the neural network significantly.

### 3.4 Affinity prediction in the absence of experimentally-determined ligand conformation using ResAtom-Score and selected scoring functions

However, the combination of $\Delta_{Vina}RF_{20}$ and ResAtom-Score would result a Pearson correlation coefficient of 0.828, which is in fact very close to the prediction capability with an experimentally determined ligand conformation. **Figure 6a** and **6b** show the RMSE and Pearson correlation coefficients predicted by different model combinations without the conformation of experimentally determined ligand. The combined use of $\Delta_{Vina}RF_{20}$ and ResAtom-Score could effectively reduce fluctuations in comparison to a single model (**Figure 6d** and **6e**).

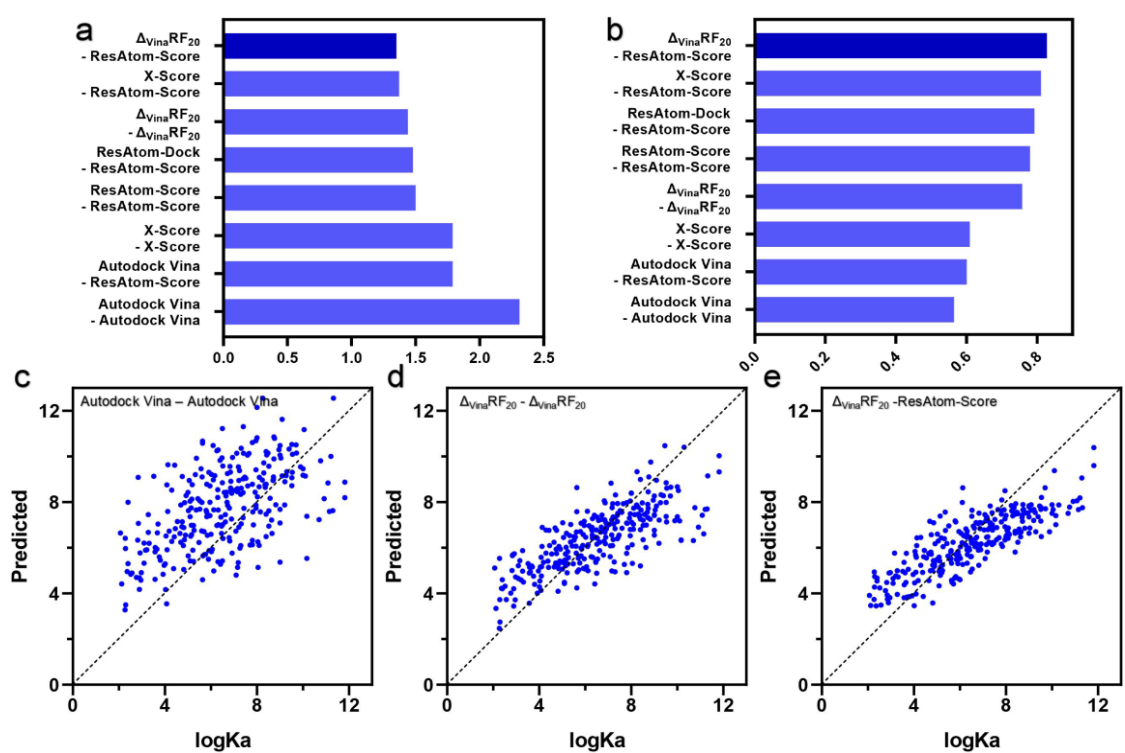

**Figure 6**. The performance of different models combined in the CASF-2016 benchmark test set without natural conformation. (a) RMSE (b) Pearson's correlation coefficient (c) Autodock Vina combined with Autodock Vina (d) $\Delta_{Vina}RF_{20}$ combined $\Delta_{Vina}RF_{20}$. (e) $\Delta_{Vina}RF_{20}$ combined with ResAtom-Score. Other results are shown in **Figures S1**.

In order to explore the reason for the improved prediction performance of VinaRF20/ResAtom-Score in the absence of experimentally determined ligand conformation, we calculated the standard deviation of their prediction values for the top three conformations scored by $\Delta_{Vina}RF_{20}$ (**Figure 7**). Compared with $\Delta_{Vina}RF_{20}$, it seems that ResAtom-Score is less affected by conformational RMSD changes. Convolutional neural networks could extract features through convolutional layers, use pooling layers to reduce complexity, and improve feature invariance (Albawi et al., 2017; O'Shea and Nash, 2015). By learning from a large PDBbind datasets, ResAtom-Score may have achieved more flexible scoring performance than other scoring models, such as $\Delta_{Vina}RF_{20}$. In the future, using a larger number of pooling layers in ResAtom-Score may further reduce the impact of slight changes in RMSD on the final scoring performance.

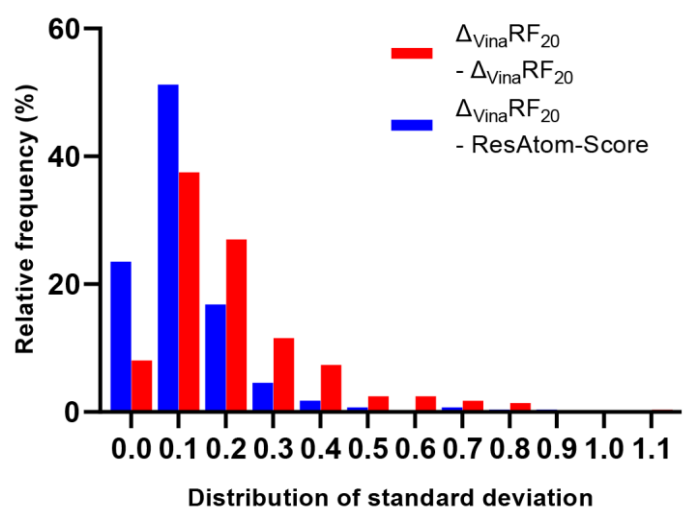

**Figure 7**. The standard deviation of the predicted value of top 3 conformations obtained by $\Delta_{Vina}RF_{20}$.

### 4. conclusion

Herein we developed a new protein-ligand affinity prediction model ResAtom-Score and a conformation screening model ResAtom-Dock using the ResNet network structure. The improved predictive ability of ResAtom system has been achieved by adding new channels and attention mechanisms in the training of PDBbind datasets, as well as the use of CosineAnnealingLR to improve the model's predictive ability. Benchmark CASF-2016 test showed that ResAtom-Score has a prediction ability of R = 0.833, superior or in par with the previously reported scoring functions in terms of affinity. We further discovered that the combination of VinaRF20/ResAtom-Score could predict protein-ligand affinity in the absence of experimentally determined ligand conformation, with an R = 0.828.

Although ResAtom performs well on the CASF2016 benchmark test set, it may only accurately predict protein-small molecule affinity. In the PDBbind database, there are only 2 594 and 973 protein-protein and protein-nucleic acid interaction data, respectively. Although there have been some reports on protein-protein interactions and protein-nucleic acid interactions, the amount of data still severely limits the performance of the model(Dai and Bailey-Kellogg, 2021; Shen et al., 2021). Despite these limitations, the development of ResAtom system may be instrumental for the development of more accurate prediction of protein-ligand affinity algorithms, and the implementation of these scoring functions for the rapid *in silico* screening for the discovery of potential drug hits.

**Acknowledgements**

This work was supported in parts by the NSFC Grant 81473124 (to Y. H.); the Chinese Ministry of Education 111 Project BP0820034 (to Y. D.) This work was also supported in part by the High Performance Computing Center of Central South University.

# Supporting Information

## ResAtom System: Protein and Ligand Affinity Prediction Model Based on Deep Learning

Yeji Wang[a], Shuo Wu[a], Yanwen Duan[a, b, c], and Yong Huang[a,c,*]

[a]Xiangya International Academy of Translational Medicine, Central South University, Changsha, Hunan, 410013, China

[b]Hunan Engineering Research Center of Combinatorial Biosynthesis and Natural Product Drug Discover, Changsha, Hunan, 410011, China

[c]National Engineering Research Center of Combinatorial Biosynthesis for Drug Discovery, Changsha, Hunan, 410011, China

*Corresponding author.

E-mail address: jonghuang@csu.edu.cn (Y. Huang)

# Table of Contents

**Table S1.** Validation performance with different learning rates and different optimizers.

| optimizer | architecture | Learning rate | R |
|---|---|---|---|
| Adam | Based on ResNet-34 | 0.01 | 0.744 |
|  |  | 0.001 | 0.754 |
| SGD |  | 0.01 | 0.709 |
|  |  | 0.001 | 0.726 |

**Table S2.** Validation performance with different architecture and different Learning rate decline strategies.

| optimizer | architecture | Learning rate | Learning rate decline strategy | R |
|---|---|---|---|---|
| Adam | Based on ResNet-18 | 0.001 | ExponentialLR | 0.722 |
| | | 0.001 | StepLR | 0.731 |
| | | 0.001 | CosineAnnealingLR | 0.739 |
| | Based on ResNet-34 | 0.001 | ExponentialLR | 0.762 |
| | | 0.001 | StepLR | 0.765 |
| | | 0.001 | CosineAnnealingLR | 0.786 |

**Table S3.** The performance of different conformation screening models.

| Model Name | TOP1 | TOP2 | TOP3 |
|---|---|---|---|
| Autodock Vina | 8.8% | 18.2% | 26.7% |
| ResAtom-Dock | 49.1% | 58.9% | 65.6% |
| X-Score | 63.5% | 74.0% | 89.4% |
| ΔVinaRF20 | 89.1% | 94.4% | 96.5% |

**Table S4.** Scoring model performance in the absence of experimentally-determined conformation.

| Screen Model | Score Model | R | RMSE | MAE |
|---|---|---|---|---|
| Autodock Vina | Autodock Vina | 0.565 | 2.31 | 1.91 |
| Autodock Vina | ResAtom-Score | 0.601 | 1.79 | 1.28 |
| X-Score | X-Score | 0.609 | 1.79 | 1.46 |
| ΔVinaRF20 | ΔVinaRF20 | 0.758 | 1.44 | 1.13 |
| ResAtom-Score | ResAtom-Score | 0.780 | 1.50 | 1.21 |
| ResAtom-Dock | ResAtom-Score | 0.792 | 1.48 | 1.20 |
| X-Score | ResAtom-Score | 0.811 | 1.37 | 1.12 |
| ΔVinaRF20 | ResAtom-Score | **0.828** | **1.35** | **1.09** |

**Figure S1.** Affinity prediction by different scoring models in the absence of experimentally-determined conformation. (a) Autodock Vina - Autodock Vina (b) Autodock Vina - ResAtom-ensemble (c) X-Score - X-Score (d) ΔVinaRF20 - ΔVinaRF20 (e) ResAtom-ensemble - ResAtom-ensemble (f) Screen Model - ResAtom-ensemble (g) X-Score - ResAtom-ensemble (h) ΔVinaRF20 - ResAtom-ensemble

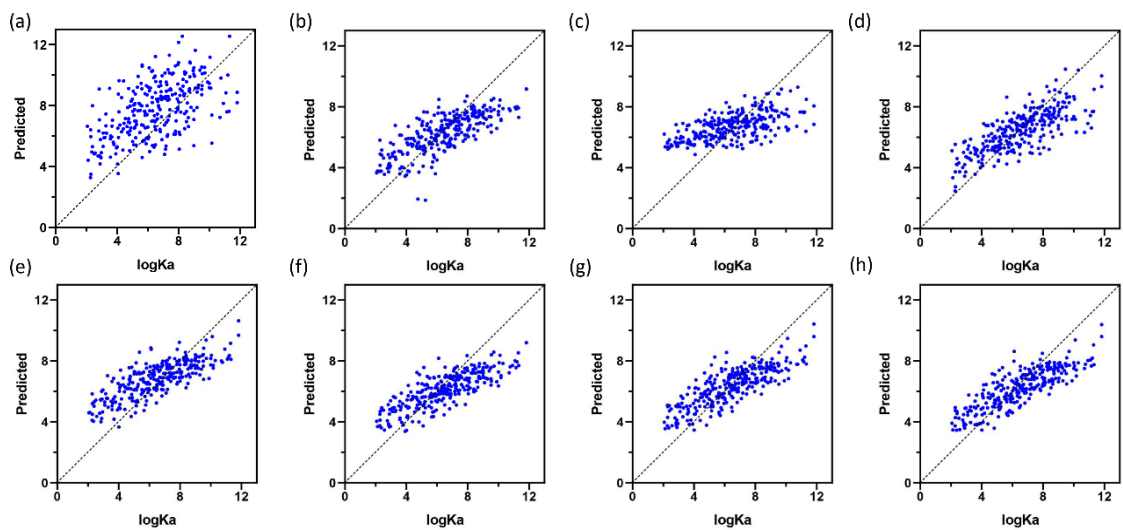

**Figure S2.** Distribution of RMSDs.

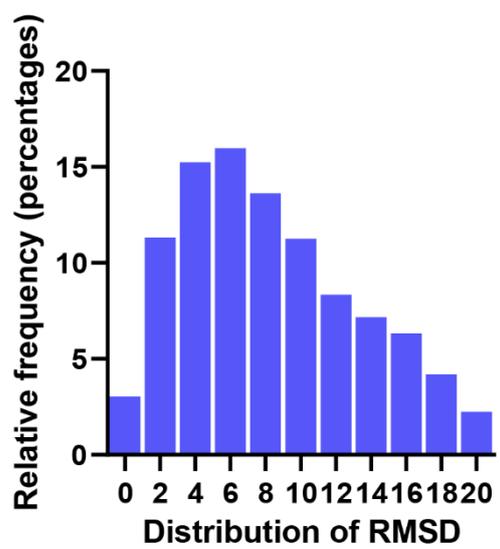